# Bright mid-infrared photoluminescence from high dislocation density epitaxial PbSe films on GaAs

Jarod Meyer[1], Aaron J. Muhowski[2], Leland Nordin[1], Eamonn Hughes[3], Brian Haidet[3], Daniel Wasserman[2], Kunal Mukherjee[1*]

[1]*Department of Materials Science and Engineering, Stanford University, CA 94306, USA*

[2]*Electrical and Computer Engineering Department, University of Texas, Austin, TX 78705, USA*

[3]*Materials Department, University of California, Santa Barbara, CA 93106, USA*

## Abstract

We report on photoluminescence in the 3–7 µm mid-wave infrared (MWIR) range from sub-100 nm strained thin films of rocksalt PbSe(001) grown on GaAs(001) substrates by molecular beam epitaxy. These bare films, grown epitaxially at temperatures below 400 °C, luminesce brightly at room temperature and have minority carrier lifetimes as long as 172 ns. The relatively long lifetimes in PbSe thin films are achievable despite threading dislocation densities exceeding $10^9$ cm$^{-2}$ arising from island growth on the nearly 8% lattice- and crystal-structure-mismatched GaAs substrate. Using quasi-continuous-wave and time-resolved photoluminescence, we show Shockley-Read-Hall recombination is slow in our high dislocation density PbSe films at room temperature, a hallmark of defect tolerance. Power-dependent photoluminescence and high injection excess carrier lifetimes at room temperature suggest that degenerate Auger recombination limits the efficiency of our films, though the Auger recombination rates are significantly lower than equivalent, III-V bulk materials and even a bit slower than expectations for bulk PbSe. Consequently, the combined effects of defect tolerance and low Auger recombination rates yield an estimated peak internal quantum efficiency of roughly 30% at room temperature, unparalleled in the MWIR for a severely lattice-mismatched thin film. We anticipate substantial opportunities for improving performance by optimizing crystal growth as well as understanding Auger processes in thin films. These results highlight the unique opportunity to harness the unusual chemical bonding in PbSe and related IV-VI semiconductors for heterogeneously integrated mid-infrared light sources constrained by tight thermal budgets in new device designs.

[*] Author to whom correspondence should be addressed: kunalm@stanford.edu

**I.  Introduction**

Narrowband incoherent and coherent mid-infrared light sources emitting in the mid-wave infrared (3−7 µm) are increasingly in demand for a range of applications such as chemical spectroscopy, healthcare, manufacturing and environmental monitoring, and free space communication.[1,2] Optoelectronic active regions in the visible and near-infrared yield devices with internal quantum efficiencies (IQE) exceeding 90%.[3] In the MWIR, however, bulk II-VI and III-V materials suffer from low room-temperature IQEs around 5%, and even highly engineered III-V superlattices are below 10%.[4–7] Interband Cascade Lasers (ICLs) have achieved high room-temperature IQEs of 64% via recycling of carriers between active stages, but they require complex band structure engineering and device design.[8,9] Although higher efficiency MWIR emitters are appealing for next generation devices, there is also substantial interest in expanding their functionality beyond simple emission. These include seamless heterogeneous integration with cost-effective substrates like Si and light manipulation through plasmonic[10] or electronic/structural phase-change engineering.[11] In this context, the mixed- ionic, metallic, and covalently bonded IV-VI (Pb,Sn)(Se,Te) rocksalt semiconductor family offers substantial advantages over III-V mid-infrared emitters. These advantages include low temperature epitaxy (<350 °C) compatible with back-end-of-line integration thermal budgets,[12] phase-change properties,[13] topologically non-trivial states,[14] smaller bulk Auger recombination coefficients,[15,16] and, potentially, a high defect tolerance.[17] Room temperature performance of PbSe- and PbTe-based mid-infrared light emitters, however, currently lags behind III-V counterparts in development.[1] Challenges include high crystal defect densities from severe lattice- and thermally-mismatched substrates, device heating due to low thermal conductivity of IV-VI materials, poor carrier confinement from incompatible barrier materials, and challenging stoichiometry and composition control in heterostructures. To develop IV-VI materials for next-generation mid-infrared light sources, we need to understand which limitations are fundamental, and which are alterable through materials engineering.

The impact of defects, especially dislocations, on the minority carrier properties of PbSe is central to the performance of heterogeneously integrated light emitting devices. Despite low Auger coefficients, attributed to simple mirrored conduction and valence band structure[18], the high background doping in PbSe (and PbTe) is reported to bring about Auger limited minority carrier lifetimes at room temperature.[19] Low-temperature Shockley-Read-Hall (SRH) recombination lifetimes, in which charge carrier de-trapping is significant, have been measured in PbSnTe via photoconductivity decay experiments to 50ns to >1us, though the limiting factors such as crystal defects, poor quality interfaces, or surfaces states have not been identified.[20] Zogg et al. characterized low-excitation recombination properties in 10 µm thick PbSe films (n-type, $2\times10^{17}$–$1\times10^{18}$ cm$^{-3}$), measuring 10–40 ns Auger-limited minority carrier lifetimes above 250 K. A number of slow traps, active at low temperatures, also extended the lifetime to several µs.[21] Klann et al. probed the Auger coefficient in 2 µm thick PbSe films (p-type, $1.6\times10^{17}$ cm$^{-3}$) in the high-injection regime, finding C = $1.1\times10^{-28}$ cm$^6$ s$^{-1}$ at 300 K.[15] Their samples have low-injection lifetimes of around 100 ns attributed to either radiative or SRH channels. In all of these studies, the specific impact of threading or misfit dislocations is not mentioned. Therefore, there is a need to clarify whether Auger or SRH limits the lifetime in high dislocation density PbSe; the recombination mechanism which limits the lifetime has significant implications for the potential for heterogeneous integration.

In this work, we study light emission, using photoluminescence (PL), from sub-100 nm epitaxial PbSe bare thin films with dislocation density exceeding $10^9$ cm$^{-2}$, grown directly on technologically-relevant GaAs(001) substrates by molecular beam epitaxy. We find these PbSe thin films emit brightly even at room temperature despite this large defect density, and exhibit SRH-limited minority carrier lifetimes at low injection. At high injection, recombination indeed becomes Auger-limited. The injection dependence of the carrier lifetime, however, suggests the Auger recombination process is degenerate and even weaker than previously reported.

## II.     Materials and Methods

Epitaxial films of PbSe were grown in a Riber Compact 21 MBE system on previously epiready oxide-desorbed and arsenic-capped GaAs(001) substrates. After desorbing the arsenic cap, the substrates were exposed to a PbSe compound flux at 400 °C for 30 s to transform the GaAs surface from a 2×4 into a 2×1 reconstruction to enable subsequent (001)-orientation cube-on-cube nucleation at lower temperatures, similar to a procedure followed for growth on InAs substrates.[22] We note that a recent report of single-orientation PbSe on GaAs(001) epitaxy was achieved without this surface treatment.[23] Sample A was grown via a single-step, "cold" nucleation and growth at 300 °C. Sample B was grown via a 2-step "hot" nucleation process where the initial nucleation was performed at 330 °C for a layer thickness below 5 nm, before cooling down to 280 °C for further growth. The hotter initial nucleation in Sample B was utilized for better orientation control, as lower temperature (001) growths tended to have small amounts of misoriented (110) nuclei visible in RHEED that are eventually overgrown. Both samples were approximately 60 nm thick and grown without extrinsic doping. Evaporation from a stoichiometric PbSe compound cell without an excess Se flux from a secondary cell typically results in n-type films in the range of $10^{17}$–$10^{18}$/cm$^3$.

The microstructures of the films were characterized using electron microscopy. Scanning transmission electron micrographs were collected at 200 kV in a TFS Talos electron microscope on focused ion beam thinned samples. Scanning electron micrographs were collected on a TFS Apreo-S microscope at 2 kV. Strain in the films was characterized using reciprocal space maps (RSM) taken in grazing-exit geometry with a PANalytical X'Pert PRO MRD system using a 1-D Pixcel Detector and Cu Kα-1 radiation.

Quasi-continuous wave (qCW) pump photoluminescence (PL) measurements were performed in a temperature-controlled cryostat with a ZnSe window, and pumped via an 808 nm laser with an output power of 1 W and a modulation frequency of 10 kHz with a 50% duty cycle. The laser was passed through a 3 μm dichroic beamsplitter and focused onto samples with an all-reflective objective. The sample PL was then focused via a reflective parabolic mirror onto a liquid nitrogen cooled MCT detector for spectrally integrated measurements, or directed to an FTIR for spectrally resolved measurements. An AR-coated (3–5 μm) Si window was used to filter out laser light, and a lock-in amplifier was used to demodulate the detector signal. A Fourier Transform Infrared Spectrometer operating in step-scan mode was used to collect the spectrally-resolved PL spectra and separate it from the thermal background radiation. The laser was defocused into an ellipse with principal radii of ~ 135 and 162 μm (spot size of 0.069 $mm^2$), and a set of neutral density filters was used to vary the incident laser power from 0.099 – 186.56 mW. Time resolved photoluminescence (TRPL) measurements were performed using a pulsed (< 1ns) 532 nm laser with a 10 kHz repetition rate and 1/e spot size of 9.73 $mm^2$. PL was collected via a parabolic mirror and focused onto a high-speed Vigo MCT detector. Samples were housed in a low-temperature cryostat, and pulse energy was controlled with neutral density filters from 0.0412 – 1.31 μJ. A LeCroy 12-bit high speed oscilloscope was used to collect and average the detected signal. The time resolution of the TRPL setup is limited by the detector response time to ≈ 5 ns.

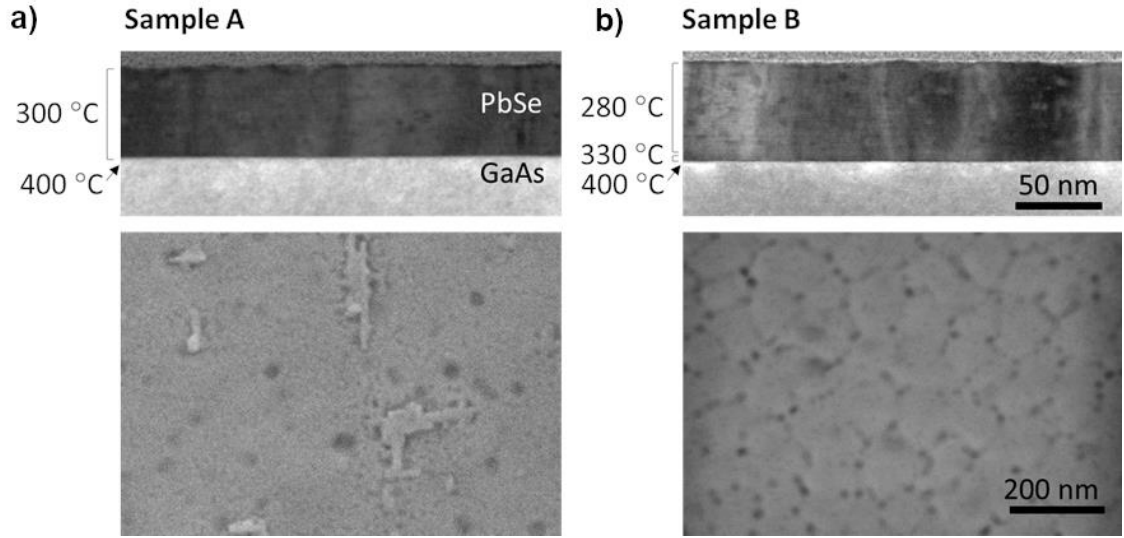

**Figure 1.** Cross-sectional STEM (top) and plan-view SEM (bottom) images of PbSe(001) on GaAs(001) in **a)** Sample A and **b)** Sample B, respectively. The temperatures to the left of the images correspond to the growth temperatures: a 400°C surface treatment followed by either growth at 300 °C, or an initial nucleation at 330 °C followed by growth at 280 °C for Samples A and B, respectively. A number of threading dislocations are seen in the STEM image for both samples. The SEM images show surfaces roughness from island coalescence artifacts.

Sample reflectivity for the qCW-PL and TRPL experiments was calculated via the transfer matrix method considering a PbSe film with air/PbSe and PbSe/GaAs substrate interfaces.

### III. Results and Discussion

#### A. Structure and strain

Figure 1a) and 1b) show cross-sectional scanning transmission electron microscope (STEM) images of Sample A and B. Both samples show a columnar grain structure of very slightly misoriented islands, with numerous threading dislocations in the PbSe film with a density exceeding $10^9/cm^2$. Note that misfit dislocations at the PbSe/GaAs interface cannot be seen at this magnification, but an 8% mismatch implies misfit dislocations spaced every few nm. Plan-view scanning electron microscope images reveal some differences in the surface morphology (Figure 1). Sample A has an overall smoother surface but shows regions that are yet to fully coalesce, whereas Sample B has a more defined grain structure with boundaries decorated with shallow pits that are likely due to threading dislocations. The microstructures of the thin films are a consequence of the highly mismatched growth, leading to very flat and shallow island nucleation followed by coalescence (Volmer-Weber growth mode); note that the RHEED pattern remains streaky across the transition from GaAs to PbSe for (001) nucleation.

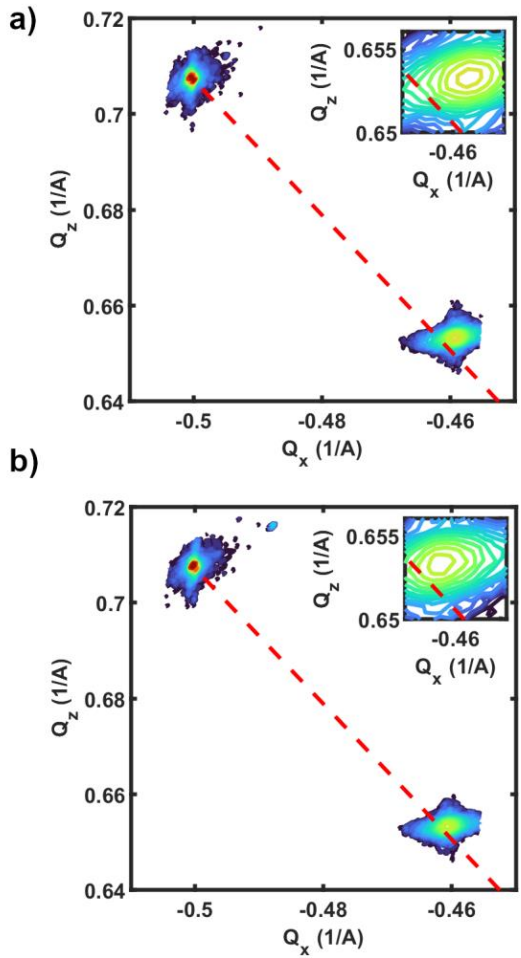

**Figure 2.** Reciprocal Space Maps of Sample A for the **a)** [224] and **b)** [$\bar{2}$24] asymmetric Bragg reflections. The full-relaxation line (red line) is also plotted, extending from the GaAs substrate peak. The inset shows the offset of the PbSe peaks due to tensile strain.

Figures 2a) and 2b) show the room temperature (224) and ($\bar{2}$24) RSMs for Sample A, respectively. From the RSMs, we measure a mean, in-plane tensile strain of 0.32% in Sample A (0.43% along [110] and 0.22% along [$\bar{1}$10]).[24] We obtain a similar result for Sample B (Fig. S1 in supplementary material). Dislocations and island coalescence essentially relax all of the nearly 8% lattice mismatch induced strain between the PbSe film and GaAs substrate during growth.[25] However, the thermal expansion coefficients for PbSe and GaAs are very dissimilar at 19.4 and $5.7 \times 10^{-6}$ K$^{-1}$, respectively, and can lead to significant tensile strain during cooldown from growth *if* dislocations are unable to glide.[26] This is indeed the case for the (001)-oriented PbSe films. These films build up significant tensile strain as dislocations experience no shear due to thermal expansion mismatch on their {100}<110> primary slip system, and the mean strain agrees with an elastic accommodation of thermal expansion mismatch from a growth temperature of 300 °C. To avoid cracking due to this strain, (001)-oriented PbSe films are necessarily thin. This is not a limitation for other orientations such as (111), however.[26]

**B. Strain-shifted luminescence and carrier recombination mechanisms**

We probe the optical properties of the 60 nm PbSe on GaAs films using qCW-PL as a function of temperature and pump power. Figures 3a) and 3b) show the temperature dependent qCW-PL spectra of Samples A and B respectively. The subwavelength films on a transparent substrate have light extraction advantages over thicker PbSe films and substrates, where the high refractive index (n = 4.8) otherwise leads to significant photon reabsorption that is especially detrimental for low IQE materials. The thin films continue to luminesce via spontaneous emission even at high injection, thus avoiding the transition to stimulated emission that is common in the IV-VI materials for micrometer thick films.[15,27] Overall, the PL spectra from both samples are nearly identical at room temperature, exhibiting FWHM of around 70 meV (2.7k$_B$T). Upon cool down, the PL intensity of Sample A drops dramatically while the FWHM only reduces slightly to

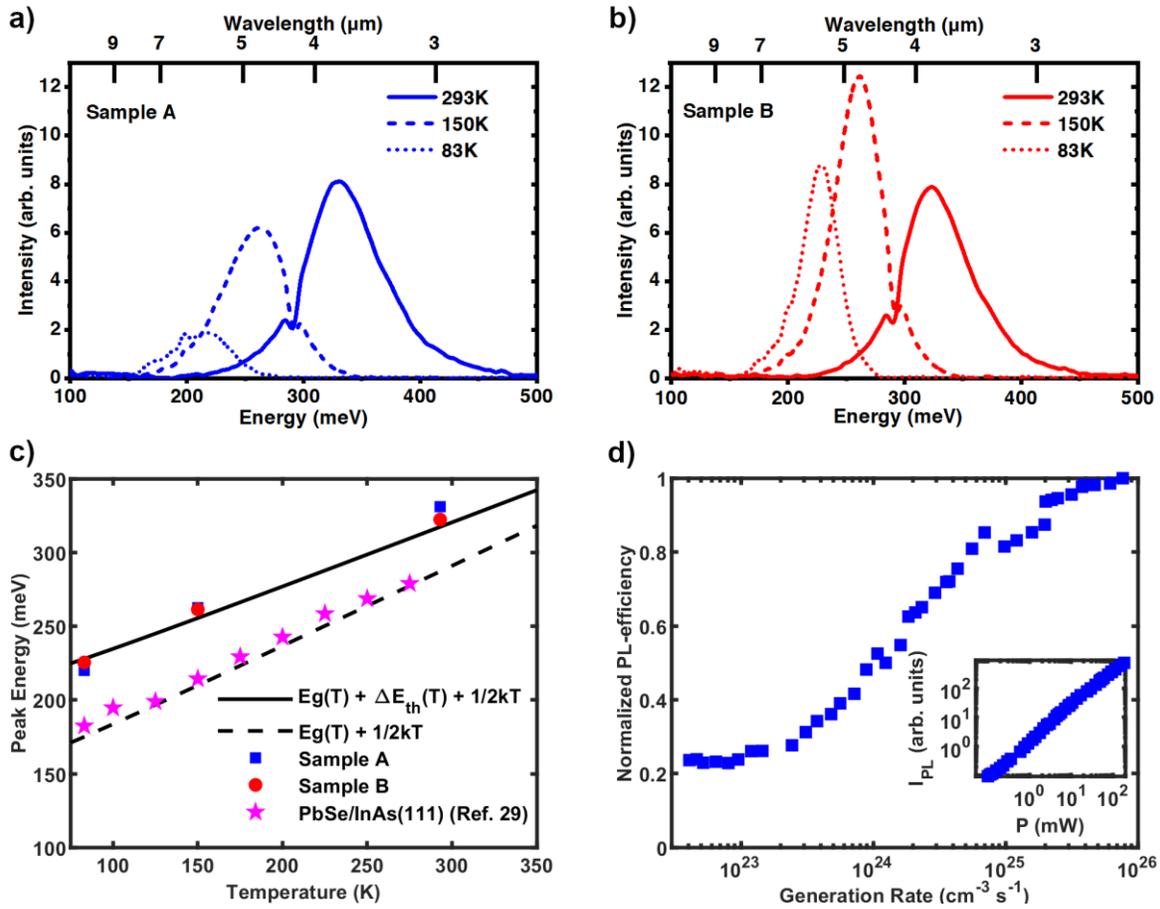

**Figure 3.** Continuous-Wave (CW) Photoluminescence spectra for **a)** Sample A and **b)** Sample B at low- and room-temperatures. **c)** Photoluminescence peak energy vs temperature for Samples A and B, along with the theoretical dependencies for thermally-strained and unstrained films (solid and dashed lines, respectively). **d)** Normalized PL-efficiency vs average Generation Rate ($G_{Avg}$) for Sample A. Plotted in the inset is the integrated photoluminescence intensity, $I_{PL}$, vs excitation power (P).

60 meV at 83K, quite broad compared to an ideal emitter ($8k_BT$ vs. an ideal $1.8k_BT$ for bulk semiconductors). The PL Intensity of Sample B, meanwhile, was brighter at 150K than at the other two temperatures measured, and was also brighter than the PL intensity of Sample A at any temperature. Sample B has a more reduced FWHM of 40 meV ($5.6k_BT$) at 83K but is still far from ideal. It appears that the initial hot nucleation at 330 °C for Sample B leads to higher material uniformity, as evidenced by the decreased FWHM of PL and high overall brightness. The slightly lower qCW-PL intensity at room temperature when compared to Sample A could be due to intrinsic (Auger) or extrinsic (SRH) recombination mechanisms thanks to variability in unintentional doping between samples which we will discuss shortly. Large unintentional background carrier concentrations can be controlled in the future using extrinsic doping.

The temperature dependence of the peak PL energy also shows a significant impact of thermal expansion-induced strain. Figure 3c) shows the peak PL energy for Samples A and B together with that from a previously reported 2 µm PbSe film on InAs (111)A, as well as the predicted bulk PbSe bandgap temperature dependence.[28,29] The (111) orientation of the thick sample permits complete relaxation of the thermal strain via dislocation glide, leading to a peak energy that agrees well with the bulk bandgap, taken here as $0.5k_BT$ below the peak PL energy. On the other hand, both Sample A and B are blue-shifted by as much as 50 meV at 83K (~30% of the bandgap). A biaxial in-plane strain for the (001) orientation deforms but preserves the degeneracy of the conduction and valence band minima of PbSe that are along <111>. The energy gap shift $\Delta E_{th}$ is calculated from the following relation[30]:

$$\Delta E_{th}(T) = \frac{2(C_{11} - C_{12})}{3C_{11}} D_{iso} \varepsilon_{th}$$
$$\varepsilon_{th} = (\alpha_{PbSe} - \alpha_{GaAs})(T_{Growth} - T),$$
(1)

where $C_{11}$ and $C_{12}$ are PbSe elastic constants, $D_{iso}$ is the isotropic deformation potential, and $\varepsilon_{th}$ is the thermal mismatch strain; we use the mean value over the anisotropic ones for the latter as an approximation. The observed peak energy agrees well with our estimate for the strain-energy increased bandgap of both samples. The recent report on MBE-synthesized PbSe on (001)-GaAs films also highlights a similar thermal-strain induced blueshift of the bandgap using absorption measurements, further corroborating our interpretation of the temperature-dependent qCW-PL spectra.[23]

The bright PL measurements motivated a more detailed characterization of carrier recombination in PbSe on GaAs at room temperature using power-dependent qCW-PL. This can reveal trends in the IQE. Generally, the integrated photoluminescence intensity for n-type material, $I_{PL}$, is proportional to the radiative recombination rate:[31]

$$I_{PL} = \chi B(\Delta n + n_0)\Delta p, \qquad (2)$$

where $\chi$ is a proportionality constant determined by details of the setup such as the excitation active volume and detector collection efficiency, B is the radiative recombination coefficient for PbSe, $\Delta n = \Delta p$ are the steady-state electron and hole densities, and $n_0$ is the electron doping concentration. Under steady-state conditions, the average generation rate, $G_{Avg}$, is then equal to the total recombination rate in the film. $G_{Avg}$ can be calculated from the laser setup and material parameters as:[32]

$$G_{Avg} = \frac{P(1-R)[1-\exp(-\alpha t_{Film})]}{A_{Spot} t_{Film} E_{Photon}}, \qquad (3)$$

where P is the laser power measured prior to hitting the sample, R is the sample reflectivity (~ 0.49), α is the PbSe Absorption coefficient (~ $2.64 \times 10^5$ cm$^{-1}$ at 808 nm[33]), $t_{Film}$ is the sample thickness (60 nm), $A_{Spot}$ is the spot size, and $E_{Photon}$ is the incident photon energy (1.53 eV). The PL-efficiency, $I_{PL}$ divided by $G_{Avg}$, is then directly proportional to the IQE at any given $G_{Avg}$, and can be calculated as:[31,34]

$$\frac{I_{PL}}{G_{Avg}} = \frac{\chi B(\Delta n + n_0)\Delta p}{G_{Avg}} = \chi \times IQE. \qquad (4)$$

Despite broader luminescence peaks and suboptimal growth initiation, we focus primarily on Sample A in this work as the minority carrier lifetime of Sample B is found to be an order of magnitude shorter at room temperature (see Section III.C), likely due to higher background doping which makes it difficult to probe the high injection regimes. Figure 3d) shows the PL-efficiency, normalized to its highest value, as a function of $G_{Avg}$ for Sample A. The inset shows the $I_{PL}$ vs P plot from which the PL-efficiency curve was derived. As stated above, the PL-efficiency (Figure 3d) is a relative measure of the IQE from which we can infer that at low $G_{Avg}$, the IQE remains constant and indicates that the low pump powers used in this study correspond to low injection conditions ($\Delta n < n_0$). Between a $G_{Avg}$ of $3 \times 10^{23}$ and $10^{24}$ cm$^{-3}$ s$^{-1}$, $\Delta n$ begins to rise above $n_0$, and IQE increases by roughly 4× from low-injection to a near peak value at a $G_{Avg}$ of $7.5 \times 10^{25}$ cm$^{-3}$ s$^{-1}$. The increase in IQE shows that SRH recombination dominates at low-injection, and radiative recombination becomes increasingly dominant over SRH recombination at higher $G_{Avg}$ since the rates of these two processes have quadratic and linear dependencies on injection, respectively. The apparent saturation of the IQE to a peak value at the highest $G_{Avg}$ is likely due to the increasing contribution of Auger recombination. Surprisingly, the IQE does not decrease with increasing $G_{Avg}$, indicating that an Auger-dominated regime is not reached. This suggests that either the SRH recombination rate is quite fast, or the Auger recombination rate is quite slow, because in both cases the peak IQE is pushed to higher $G_{Avg}$. We show subsequently that a long minority carrier lifetime in Sample A indicates that the latter case.

## C. Low and high injection carrier lifetimes at room temperature

We use TRPL measurements to directly measure recombination lifetimes and their dependencies on carrier injection and temperature. For the 532 nm pump light used, the absorption coefficient of PbSe is roughly $6.50 \times 10^5$ cm$^{-1}$,[33] which corresponds to ~ 98% of the transmitted pump light being absorbed within the 60 nm films. Photo-excited carriers then rapidly diffuse to yield a uniform initial carrier concentration, Δn, that is linear with the pulse energy. Δn was calculated for the TRPL experiment via:[35]

$$\Delta n = 0.683 \times \frac{\eta P_{Pulse}(1-R)[1-\exp(-\alpha t_{film})]}{A_{Spot} t_{film} E_{photon}}. \tag{5}$$

Where $P_{Pulse}$ is the total energy per pulse, and R was ~ 0.42. The carrier multiplication efficiency, $\eta \approx 1.25$, was included in the expression since on average 2.33 eV photons generate more than one electron-hole pair in bulk PbSe.[36]

Figure 4a) shows the room temperature, injection-dependent TRPL decay for Sample A. From the single-exponential decay regions, low-injection minority carrier lifetimes, $\tau_{mc}$, were found to be 172 ± 18 ns at room temperature despite threading dislocation densities greater than $10^9$ cm$^{-2}$ and suggest a high degree of dislocation tolerance compared to III-V and HgCdTe emitters.[37,38] Promisingly, the PL-efficiency curve for Sample A in Figure 3d) suggests that the $\tau_{mc}$ of 172 ns is still SRH-limited, and further advances in material quality should lead to even longer lifetimes. We measured a minority carrier lifetime of 16 ± 4 ns at room temperature for Sample B (Fig. S2) which we attribute to higher background doping (see supplemental material).

We estimate the recombination lifetime at high injection by noting the luminescence decay right after the very short laser pulse of known energy; the upper bound of this analysis is set by the instrument time resolution of 5 ns. The initial (or instantaneous) lifetime, $\tau_{initial}$, at the time of the laser pulse injection of excess carriers Δn was calculated as:[39]

$$\frac{1}{\tau_{initial}} = -\frac{1}{\Delta n}\frac{d\Delta n}{dt} = \frac{d\Delta n}{dS}\frac{dS}{dt}, \tag{6}$$

where S is the maximum TRPL signal. dΔn/dS was determined by fitting the dependence of maximum TRPL signal vs Δn, and dS/dt was derived from the decay plots in Figure 4a). Figure 4b) plots $\tau_{initial}$ vs Δn for Sample A at room temperature, with the inset showing the dependence of $\Delta n$ on S. $\tau_{initial}$ decreases from roughly 74 ns at $3.6 \times 10^{17}$ cm$^{-3}$ injection, to 9 ns at $2.9 \times 10^{18}$ cm$^{-3}$ injection. For comparison with these high injection lifetimes, we also plot the intrinsic Auger lifetime assuming a constant coefficient of $1.1 \times 10^{-28}$ cm$^6$ s$^{-1}$ as $\tau_{Auger,Constant}$, along with the intrinsic radiative lifetime, $\tau_{rad}$, as derived from the literature for PbSe by Zogg et al.[21] The measured lifetimes are clearly lower than $\tau_{rad}$ but track it closely. Calculating IQE as the ratio of $\tau_{initial}/\tau_{rad}$ at each point yields a peak IQE of roughly 30%. The extracted IQE values are naturally sensitive to the value of the PbSe radiative recombination coefficient B, which from Ref. 21 was calculated to be roughly $1.2 \times 10^{-11}$ cm$^3$ s$^{-1}$, while Zhang obtains a similar, but smaller, value of $8 \times 10^{-12}$ cm$^3$ s$^{-1}$ at 300 K from first principles density functional theory calculations.[40] Conservatively, the slower recombination rate value still yields a peak IQE > 20% for Sample A at room temperature, over 50% higher than reported MWIR III-V superlattice and HgCdTe IQEs at room temperature grown on lattice-matched substrates.[4–7] Finally, we note that the near-constant IQE in high-injection also matches well with the saturation of the PL-efficiency curve at high $G_{Avg}$ in Figure 3d). This is expected since an effective generation rate in the TRPL experiment, calculated by dividing Δn by $\tau_{initial}$, is in the range of $5.0 \times 10^{24} – 3.3 \times 10^{26}$ cm$^3$ s$^{-1}$. This overlaps with the higher generation rates probed in the qCW-PL experiment, where IQE was not changing as significantly with $G_{Avg}$.

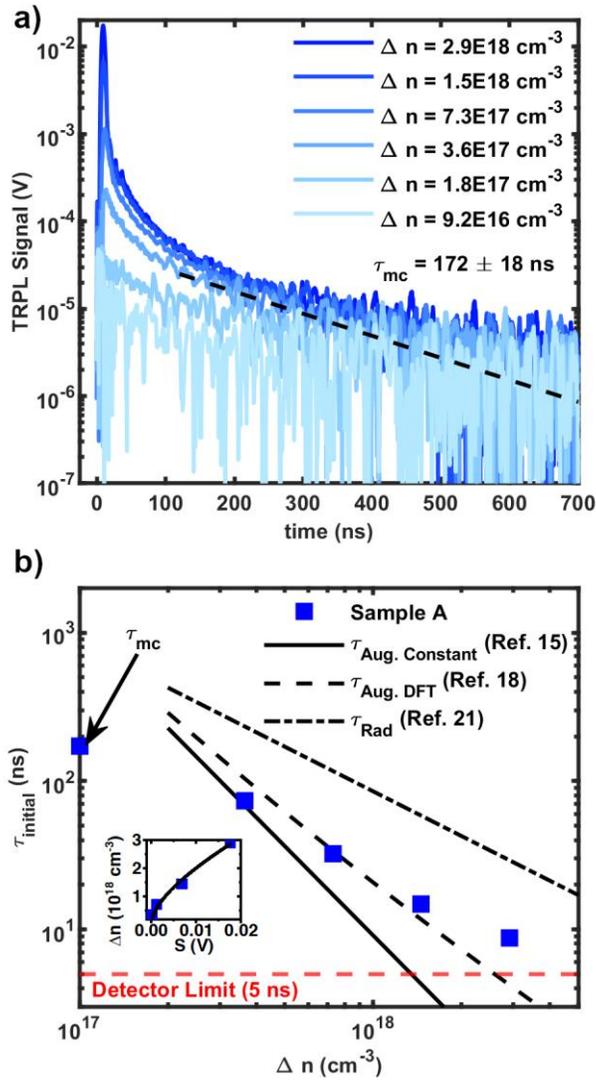

**Figure 4. a)** Injection-dependent, time-resolved photoluminescence decay for Sample A at room temperature. The black dashed line shows the low-injection lifetime, $\tau_{mc}$, as determined from the single-exponential portion of the decay. **b)** Instantaneous lifetime $\tau_{initial}$ vs injection density, $\Delta n$, for Sample A, plotted alongside the intrinsic Auger lifetime for a constant Auger coefficient of $C_n + C_p = 1.1 \times 10^{-28}$ cm$^6$ s$^{-1}$, $\tau_{Aug.\ Constant}$ (solid), the intrinsic Auger lifetime as recently computed in the literature by DFT, $\tau_{Aug.\ DFT}$ (dashed), and the intrinsic radiative lifetime for PbSe from literature, $\tau_{Rad}$ (dot-dash). The inset shows the response curve of $\Delta n$ vs the max TRPL signal, S.

It is clear that Sample A is not radiative-limited in high injection from our estimates of sub-unity IQEs, and thus Auger recombination is clearly responsible for the saturation in IQE observed in qCW-PL. Nevertheless, the measured lifetimes are still longer than the experimental (constant) Auger coefficient in literature of $1.1 \times 10^{-28}$ cm$^6$ s$^{-1}$ would suggest, and appear to have a different functional form. While the Auger-recombination lifetimes for conventional CCCH and CHHH processes go as $\Delta n^{-2}$, the measured lifetimes for Sample A have a weaker dependence. This in itself is not unexpected for Auger recombination in high-injection; Haug et al. have previously shown that in the limit of degeneracy the lifetime dependence of Auger recombination on Δn weakens from $\Delta n^{-2}$ to $\Delta n^{-1}$.[41] This phenomena has been observed before in both bulk InSb and III-V superlattices.[42,43] Recently, Zhang et al. investigated Auger recombination in PbSe from first principles using density functional theory calculations. They found that indirect, phonon-assisted Auger recombination was dominant in PbSe rather than conventional Auger, but still predicted a total Auger coefficient of $1.1 \times 10^{-28}$ cm$^6$ s$^{-1}$ at room-temperature that decreased with injection due to degeneracy effects.[18] We also plot this predicted intrinsic Auger lifetime including degeneracy, $\tau_{\text{Aug. DFT}}$, in Figure 4c). A better match to the experimental lifetime dependence on $\Delta n$ is found, considering experimental uncertainties. $\tau_{\text{Aug. DFT}}$, however, is still shorter compared to our lifetime measurement. In fact, the first principles Auger coefficient calculation would suggest that IQE should decrease past a $G_{\text{Avg}}$ of $5 \times 10^{-24}$ cm$^{-3}$ s$^{-1}$ which disagrees with the observed increase in IQE up to a $G_{\text{Avg}}$ over an order of magnitude higher in the qCW-PL experiment (Figure 3d). These results suggest that the high-injection Auger recombination rate in Sample A is both degenerate, and slower than previously reported, which bodes well for the development of MWIR light emitters. At this point, we acknowledge that the results for these 60 nm thin films may indeed not be representative of bulk behavior. Still, it is clear that the Auger recombination rate for the PbSe film on GaAs is more than two orders of magnitude lower than the rates found for bulk, III-V emitters in the MWIR.[44]

**D. Temperature dependence of minority carrier lifetime**

In addition to high injection Auger recombination, we may also understand the unusual decrease in luminescence intensity with decreasing temperature seen in Sample A (Fig. 3a) with TRPL. The temperature dependence of $\tau_{\text{mc}}$ of Sample A is shown in Figure 5. $\tau_{\text{mc}}$ decreases from SRH-limited 172 ns to about 20 ns as Sample A is cooled from 300 to 120 K; below 120K the lifetime becomes shorter than the time resolution of the experimental setup. The decrease in $\tau_{\text{mc}}$ for Sample A is clearly not due to the low-temperature enhancement of radiative recombination since the qCW-PL intensity is lower at cryogenic temperatures for Sample A. Furthermore, a typical T$^{-3/2}$ dependence of the radiative rate would only yield a 4× decrease of the lifetime from 300–120K, compared to the ~ 8× decrease observed. Auger recombination also should not be responsible for a shorter lifetime at lower temperatures. Direct Auger recombination is a thermally-activated process, while phonon-assisted Auger recombination, the recently-predicted dominant Auger mechanism for PbSe, is only weakly temperature-dependent.[18] Given that Sample A is SRH-limited in low-injection at room temperature, we expect the Auger contribution to $\tau_{mc}$ to be negligible at lower temperatures. Thus, we investigate this lifetime trend assuming SRH-dominated recombination also at low temperatures. The SRH lifetime, $\tau_{SRH}$, was extracted from $\tau_{mc}$ via:

$$\tau_{SRH} = \left(\frac{1}{\tau_{mc}} - \frac{1}{\tau_{Rad}}\right)^{-1}, \tag{7}$$

where $\tau_{Rad}$ is the radiative lifetime as specified by Zogg et al.[21] Although SRH recombination lifetimes are typically longer at low temperatures, they can get faster with decreasing temperature in temperature ranges for which the Fermi level lies below the trap energy level. Assuming a single energy level trap, Rogalski et al. model τ$_{SRH}$ for this scenario as:[45]

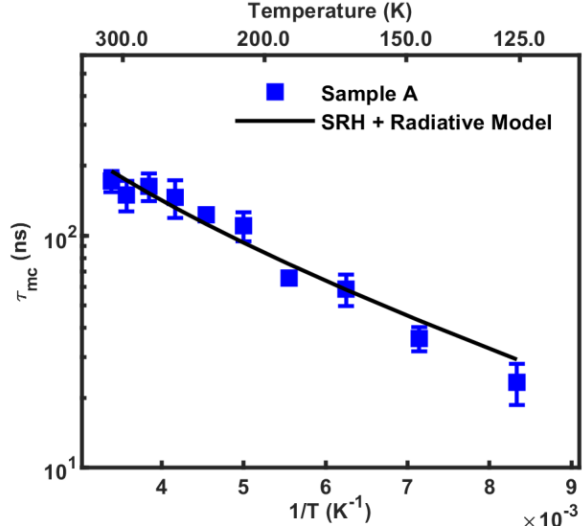

**Figure 5.** Low-injection lifetime, $\tau_{mc}$, vs temperature for Sample A. The solid line is a fit of $\tau_{mc}$ to SRH theory including radiative recombination, which yields a trap energy $E_t$ of 18.8, and a $N_t\sigma_p$ product of 0.38 cm$^{-1}$ for the relevant defect.

$$\tau_{SRH} = \frac{N_c}{\sigma_p v_{th} N_t n_0} \exp(\frac{-E_t}{kT}), \tag{8}$$

where $N_c$ is the conduction band density of states, $E_t$ is the trap energy level, $\sigma_p$ is the hole capture coefficient for the defect, $v_{th}$ is the hole thermal velocity given as $\sqrt{\frac{8k_B T}{\pi m_h^*}}$, and $N_t$ is the trap concentration. A background carrier concentration of $n_0 = 10^{17}$ cm$^{-3}$ was assumed for the fitting; reasonable given both typical unintentional doping values measured for IV-VI epitaxial layers from compound PbSe flux[46] and based on an estimate of the doping as $n_0 = G_{Avg}\tau_{mc}$, where $G_{Avg} = 6\times10^{23}$ cm$^{-3}$ s$^{-1}$ is roughly when the PL-efficiency switches from low to high-injection behavior (Figure 3d). The solid black line in Figure 5 shows the fit of τ$_{mc}$ to the SRH + Radiative model, from which we extract an $E_t$ of 18.8 meV below the conduction band, and an $N_t\sigma_p$ product of 0.38 cm$^{-1}$ (for example: $\sigma_p$ of only $4\times10^{-17}$ cm$^2$ for a trap density of $10^{16}$ cm$^{-3}$). SRH-limited recombination at low temperatures has been reported previously in the IV-VI's for PbSnTe diodes,[47] and such shallow energy trap levels are not uncommon in the MWIR; HgCdTe trap levels have been found with energies less than 11 meV from band edges in high quality samples.[48] It is important to note that this analysis assumes constant capture coefficients and trap energy levels. Capture coefficients that increase with decreasing temperature have been measured before for some impurity levels in silicon.[49,50] The large modification of the band gap due to thermal expansion-induced strain, up to a 30% increase in $E_g$ at 83K, could also deepen SRH trap energy levels with respect to the band edges. Both of these effects would also lead to increases in SRH recombination rates at cryogenic temperatures, beyond the model used for Figure 5.

Overall, it is remarkable that the energy level of the dominant traps in PbSe heterogeneously integrated on GaAs (caused by dislocations or otherwise) are quite close to the band edge, resulting in them becoming more benign at higher temperatures where the application space is widest. We do not clarify mechanisms by which minority carriers in PbSe are tolerant to dislocations since we have no discriminating experiments in this work to suggest one way or the other. Other authors have suggested the role of mixed-bonding giving rise to high dielectric constants that screen crystal defects, or the order or nature of atomic orbitals that contribute to the conduction and valence bands including lone pair effects.[51] Even details of the carrier capture pathway are important since a trap energy level in the gap does not necessarily mean faster recombination.[52] These are now common themes in discussions[53] on defect-tolerant, earth-abundant, solar-cell materials and the lead halide perovskites, which share many of these traits with PbSe.

**E. Luminescence from PbSe on alternate III-V substrates**

We have shown that minority carrier recombination in PbSe on GaAs films is SRH-limited in low-injection. Thus, growing on more lattice-matched substrates should, in theory, lead to lower threading and misfit dislocation densities and higher IQEs. To study this, growths on the more lattice-matched substrates of InAs (1.1% mismatch) and GaSb (0.5% mismatch) were attempted. Details of growths on these substrates have been reported previously, although we issue the caveat that direct PbSe growth on GaSb remains challenging and the sample may have interface reactions or multiple crystallite orientations not characterized in the present work.[22] Figure 6 compares qCW-PL from 60–80 nm PbSe films grown on InAs(001) and GaSb(001) substrates at 83K. Unfortunately, PL intensity from both films was found to be significantly weaker than the films grown on GaAs(001) substrates. Other carrier loss mechanisms, more important than dislocations, are clearly affecting the luminescence efficiency. In this case, we hypothesize brighter luminescence on GaAs is due to the more favorable band alignments brought on by growing the narrow-gap PbSe on a wide-gap GaAs substrate which may permit a type-I band offset to confine electrons and holes in the PbSe film. For InAs and GaSb, the band alignment with PbSe might be type-II, or even type-III (broken gap), as has been recently reported for PbTe/InSb,[54] leading to rapid transfer of one or both carrier types out of PbSe and a quenching of luminescence. Thankfully, carrier confinement is not a fundamental limitation as the use of AlSb-containing barriers, thicker PbSe films, or PbSnSe heterostructures become useful once key SRH-controlling material defects are identified.

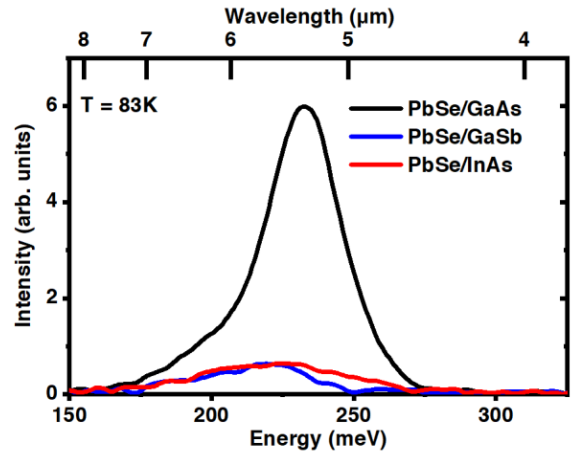

**Figure 6.** qCW-Photoluminescence Spectra at 83K for (001)-PbSe films of 70 – 100 nm thickness grown on GaAs (black), GaSb (blue), and InAs (red) (001) substrates, respectively. Despite having an 8% lattice mismatch to its substrate, the PbSe film on GaAs has a significantly stronger photoluminescence intensity than the more lattice-matched samples.

## IV.    Conclusions

In summary, the photoluminescence characteristics of thin, epitaxial films of (001)-oriented PbSe on GaAs show long, room-temperature carrier lifetimes despite high dislocation densities, comparing favorably with the popular InAs-InSb and HgCdTe systems. The PbSe films are SRH-limited at low injection and provide further opportunities to improve the IQE towards intrinsic limits. An initial nucleation temperature of 330 °C led to a higher material uniformity, as evidenced by the decreased FWHM of PL with respect to the sample grown without it but more work is needed to understand this heterovalent interface. Under high injection, the PbSe films appear limited by a degenerate Auger recombination process with a rate much lower than for equivalent III-V bulk materials and even that of bulk PbSe. Overall, due to a combined effect of the lower Auger recombination rate and tolerance to defects, we estimate a peak IQE of ~ 30% at room temperature for MWIR emission. We also show that thermally induced strain through heterogeneous integration is a viable method, beyond composition, to introduce biaxial strain in the study of IV-VI materials and for tuning light emission and absorption across the whole MWIR range. Looking ahead, we need to understand why PbSe and related materials are defect tolerant and clarify the nature of Auger recombination in IV-VI materials and the impact of strain or quantum confinement on its magnitude. Together with the attractively low synthesis temperatures for epitaxy, IV-VI materials may have serious potential as efficient light emitters for heterogeneously integrated back-end-of-line applications with significant implications for materials and device development activities.

**Supplementary Material**

See supplementary material for x-ray diffraction RSMs and time-resolved photoluminescence of sample B

**Acknowledgements**

This work was supported through National Science Foundation (NSF) CAREER award under grant No. DMR-1945321 and NSF Materials Research Science and Engineering Center (MRSEC) at UC Santa Barbara under grant No. NSF DMR-1720256 (Seed program). L.N. and D.W. acknowledge support from the National Science Foundation under grant No. ECCS-1926187. We also acknowledge the use of shared facilities of the NSF MRSEC at UC Santa Barbara, grant No. NSF DMR-1720256. B.B.H. acknowledges support from the NSF Graduate Research Fellowship under grant No. 1650114. We thank Xie Zhang and Chris Van de Walle for helpful discussions on radiative and Auger recombination in PbSe. We also would like to acknowledge Kurt Olsson and John English for their MBE expertise and support.

**Author Declarations**

The authors have no conflicts to disclose.

**Data Availability**

The data that support the findings of this study are available from the corresponding author upon reasonable request.

# Supplementary Information

**Bright mid-infrared photoluminescence from high dislocation density epitaxial PbSe films on GaAs**

Jarod Meyer[1], Aaron J. Muhowski[2], Leland Nordin[1], Eamonn Hughes[3], Brian Haidet[3], Daniel Wasserman[2], Kunal Mukherjee[1]

[1]*Department of Materials Science and Engineering, Stanford University, CA 94306, USA*
[2]*Electrical and Computer Engineering Department, University of Texas, Austin, TX 78705, USA*
[3]*Materials Department, University of California, Santa Barbara, CA 93106, USA*

## 1. Reciprocal Space Mapping of Sample B

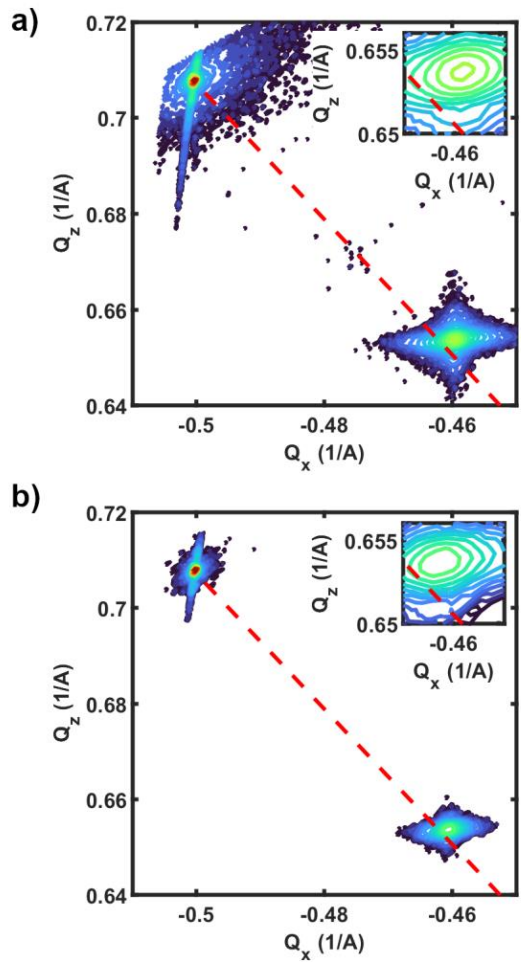

**Figure S 1.** Reciprocal Space Maps of Sample B for the **a)** [224] and **b)** [$\bar{2}$24] asymmetric Bragg reflections. The full-relaxation line (red line) is also plotted, extending from the GaAs substrate peak. The inset shows the offset of the PbSe peaks due to tensile strain.

Figure S1a) and b) show the (224) and (-224) Reciprocal Space Maps for Sample B, respectively.

From the peak positions in $Q_x$ and $Q_z$ space, the in-plane and out-of-plane lattice constants were calculated to yield the in-plane tensile strain, assuming a Poisson's ratio of 0.31 for PbSe.[23,30] Thus, a mean, in-plane tensile strain of 0.31% was measured for Sample B (0.41% along [110] and 0.21% along [$\bar{1}$10]). The measured tensile strains for Sample B are near-identical to those measured for Sample A (0.43% and 0.22%, respectively for [110] and [$\bar{1}$10]) as expected.

2. **Time-resolved Photoluminescence of Sample B**

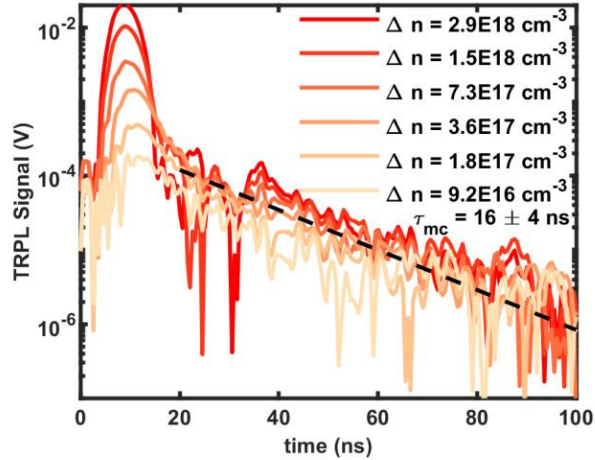

**Figure S 2.** Time-resolved Photoluminescence Decay for Sample B as a function of injection level. From the single-exponential decay regions, a minority carrier lifetime, $\tau_{mc}$, of $16 \pm 4$ ns was determined.

Figure S2 shows the injection-dependent, time-resolved photoluminescence decay traces for Sample B. Compared to Sample A, the maximum TRPL signal at lower injection was higher for Sample B despite a shorter $\tau_{mc}$ of ~ 16 ns. The TRPL signal is related to injection level via:[31]

$$S \propto B(n_0 + \Delta n)\Delta p, \quad (1)$$

where B is the radiative recombination coefficient, $\Delta n = \Delta p$ are the injected electron and hole densities, and $n_0$ is the electron doping concentration. Under low injection, $n_0 > \Delta n$, and S is essentially proportionate to $Bn_0$. Assuming the B coefficient is identical between Samples A and B, which are both PbSe films of identical thickness, then the higher S at low injection in Sample B can be attributed to a higher $n_0$ than Sample A.